%% file: tc.tex
\documentclass[twocolumn, secnumarabic, amssymb, nobibnotes, aps, prd]{revtex4-1}

\usepackage{braket}
\usepackage{graphicx}
\usepackage{graphicx}
\usepackage{mathtools}
\usepackage{amssymb}
\usepackage{amsthm}
\usepackage{amsmath}
\usepackage{subcaption} % 用于创建子图
\usepackage{siunitx}
\usepackage{booktabs}
\usepackage{float}
\usepackage{flushend} 
\usepackage[table]{xcolor}
\usepackage{placeins}
\usepackage{hyperref}
\begin{document}	
	\title{Strongly Coupled Continuous Time Crystal }%
	\author{Ximo Wang, Qiwei Han, Zhenqi Bai, Hongyan Fan, Yichi Zhang$^{\dagger }$}%
	\email[Contact author:]{zhangyichi@sxu.edu.cn}
	\affiliation{ College of Physics and Electronic Engineering, Shanxi University, 030006 Taiyuan, People’s Republic of China\\
		Collaborative Innovation Center of Extreme Optics, Shanxi University, Taiyuan, Shanxi 030006, People’s Republic of China}
	\begin{abstract}
		Time crystals are classified into discrete time crystals and continuous time crystals based on whether they spontaneously break time-translation symmetry. Continuous-time crystals do not require external driving. By introducing AdS/CFT duality to time crystals, we derive their thermodynamic limit and find that in strongly correlated many-body systems such as a 3D optical lattice (ions or tweezer in supplemental materials), cooperative many-body tunneling enables time crystals to oscillate spontaneously. In strongly correlated quantum systems driven by many-body cooperative tunneling, we discover a universal scaling law governing the time-crystalline phase transition at a critical temperature. 
	\end{abstract}
	\maketitle
	\textit{Introduction} The concept of time crystals, first proposed by Wilczek \cite{1,2,3}, represents a novel class of quantum phases characterized by the spontaneous breaking of time-translation symmetry. In contrast to spatial crystals, which exhibit long-range periodicity in space, time crystals exhibit stable periodic dynamics in time, even in the absence of external driving. Since their theoretical inception, time crystals have inspired extensive exploration across quantum many-body physics, non-equilibrium dynamics and quantum information science.
	
	The first realizations of time crystals were discrete time crystals (DTCs) \cite{4}, which break discrete time-translation symmetry under periodic driving. These were experimentally demonstrated in diverse platforms such as trapped ions \cite{5,6}, dipolar trap \cite{7,8}, Many-body localization \cite{9}, and Floquet-engineered cold atoms \cite{10,11,12,13,14,15,16} and so on \cite{17,18,19}. Discrete time-crystalline order arises from many-body localization, prethermalization, or disorder-protected synchronization mechanisms that stabilize periodic subharmonic oscillations. More recently, continuous time crystals (CTCs), which break continuous time-translation symmetry in undriven systems, have attracted growing interest due to their connection to intrinsic many-body dynamics \cite{20,21,22}. Experimental signatures of CTCs have been observed in cavity QED setups \cite{23,24}, Rydberg-dressed systems \cite{25}, and electron–nuclear spin systems \cite{26,27}. Continuous time crystals have also found applications in materials science and quantum sensing \cite{28,29}. However, a general theoretical framework for continuous time-crystalline order in strongly correlated systems remains an open challenge.
	
	In parallel, the AdS/CFT \cite{30,31,32,33,34,35} correspondence has emerged as a powerful tool in understanding quantum field theories via gravitational duals. Originally formulated in string theory, the correspondence has found wide applications in modeling strongly correlated condensed matter systems, such as holographic superconductors, non-Fermi liquids, and quantum critical transport. In this context, oscillatory scalar fields in AdS-Reissner-Nordström (RN-AdS) \cite{36,37} black hole backgrounds are known to exhibit quasinormal modes (QNMs) \cite{38,39,40,41} that naturally correspond to dissipative dynamics in the boundary theory. These holographic oscillations provide a gravitationally encoded mechanism for emergent long-lived coherence, suggesting a potential route to modeling time crystals.
	
	In this letter, we propose a holographically dual model of continuous time crystals by embedding a BEC in a three-dimensional optical lattice and coupling its collective quantum dynamics to the QNM spectrum of a charged RN-AdS black hole \cite{42,43,44,45,46} and Kerr black hole. We demonstrate that cooperative many-body tunneling can give rise to spontaneous time-translation symmetry breaking in the absence of external periodic forcing. By establishing a duality between the oscillation frequency $\omega$ and the black hole electric potential $\Phi(r_h)$, we obtain a synchronization condition $\omega=q\Phi(r_h)$ that links the boundary time crystal dynamics to bulk horizon properties. Furthermore, we derive a universal expression for the critical temperature $T_c$, which marks the onset of time-crystalline order, and depends explicitly on experimentally tunable quantities such as lattice depth, interaction strength, condensate density, and synthetic charge.
	
	We reveal the phase transition between superfluid \cite{47,48}, Mott insulator \cite{49}, and time-crystalline phases. Our approach provides a unified picture of continuous time crystals grounded in holographic duality, and suggests experimental paths for realizing such phases in cold-atom systems (BECs in optical lattice) \cite{50,51,52,53,54,55,56,57,58,59,60,61,62,63}. Importantly, the model connects emergent coherent oscillations to black hole horizon physics, offering insights into symmetry breaking, dissipationless dynamics, and critical phenomena in strongly correlated quantum matter. To support our analytical results, we construct a phase diagram.

	\textit{Holographic time crystal} Treating time crystal as a quantum field, and it reads
	\begin{equation}
		\psi(r,t) =\psi(r)e^{-i\omega t}.
	\end{equation}
	We may regard this as a quantum field on the CFT boundary, which should then correspond to an oscillating solution in the dual AdS space, likely resembling an oscillating charged black hole \cite{42,43,44,45,46} informed to offset the dissipation of quasi-regular films to satisfy the oscillation form required by time crystals. The critical temperature of the black hole is equivalent to the critical temperature of the time crystal. The action reads 
	\begin{equation}
		\begin{split}
			S=&\int D^{d+1}x\sqrt{-g} [\frac{1}{16\pi G}(R-2\Lambda+F_{\mu \nu }F^{\mu  \nu } )\\
			&\quad- \left | \partial _{\mu }\psi -iqA_{\mu }\psi  \right |^2 -V(\left |\psi   \right | ) ],
		\end{split}
	\end{equation}
	where d is space-time dimension, $g$ is negative curvature gauge, $\Lambda$ is cosmological constant, and $V$ is a scalar field. $V\left ( \left |  \psi \right |  \right )=m^2\left | \psi  \right |  ^2+\lambda\left | \psi  \right |^4 $. AdS radius $L$ satisfies $\Lambda=-\frac{d(d-1)}{2L^2} $. The system satisfies the Einstein field equations on the AdS side and the scalar field equations on the CFT boundary
	\begin{align}
		&R_{\mu \nu }-\frac{1}{2}g_{\mu \nu }R+\Lambda g_{\mu \nu } =8\pi GT_ {\mu \nu },\\
		& (\nabla_\mu-iqA_\mu ) ^2\psi -m^2\psi =\lambda\left | \psi  \right | ^2\psi ,
	\end{align}
	
		where $T_ {\mu \nu }$ is the energy-momentum tensor. The background metric is a static, spherically symmetric black hole
	\begin{equation}
			\mathrm{d}s^2=-f(r)\mathrm{d}t ^2+\frac{\mathrm{d}r^2 }{f(r)}+r^2\mathrm{d}\Omega _{d-1}^2.
	\end{equation}
	It is straightforward to derive a differential equation for the field
	\begin{equation}
		f \psi_0'' + ( f' + \frac{3f}{r}) \psi_0' + \left( \frac{(\omega-q\Phi)^2}{f}- m^2 \right) \psi_0 - \lambda \psi_0^3 = 0.
	\end{equation}
	For the following calculations, we will use three spatial dimensions (i.e., $d=4$ spacetime dimensions) as our working example. The Schwarzschild-AdS black hole metric is given by
	\begin{equation}
		f(r) = 1 + \frac{r^2}{L^2} - \frac{\mu}{r}+\frac{Q^2}{ r^2}, \quad \mu = \frac{8\pi G M}{3},  
	\end{equation}
	where $L$ is the AdS radius, $Q$ is the electric charge of a black hole. $M$ is the black hole mass, and $\mu$ parameterize the gravitational potential. $\Phi=\frac{Q}{4\pi }\frac{1}{r}$ is the electric field. The term $r^2/L^2$ dominates at large $r$, reflects AdS asymptotics, while $\mu/r^2$ encodes the black hole’s gravitational influence. With the simultaneous condition $\omega = q \cdot \left( -\left. \frac{\partial \Phi}{\partial r} \right|_{r_h} r_h \right) = q \cdot \frac{Q}{4\pi r_h}$, $T_c$ is determined by the Eq.5 when $\left. \frac{d}{d\omega} \text{Im}[\omega_{\text{bulk}}] \right|_{\omega=0} = 0$.
	And we obtain $T_c$
	\begin{equation}
		T_c=\frac{1}{4\pi}(\frac{3r_h}{L^2}-\frac{Q^2}{{r_h}^3}),
	\end{equation}(as shown in supplemental materials).

	Within the AdS/CFT duality, the spontaneous oscillation frequency $\omega_n$ of strongly correlated time crystals exhibits rigorous correspondence with quasinormal mode (QNM) frequencies $\omega_\mathrm{QNM}$ of charged RN-AdS black holes: $\omega_n = \mathrm{Re}[\omega_\mathrm {QNM}]$. The synchronization condition $\omega_\mathrm{QNM} = q\Phi(r_h)$ locks the boundary-bulk dynamics. Crucially, at the critical temperature $T_c$, the vanishing scalar condensate amplitude drives a universal scaling $\mathrm{Im}[\omega_\mathrm{QNM}] \propto-(T - T_c)^{1/2}$, with exact dissipationless oscillation emerging at $T_c$ where $\mathrm{Im}[\omega_\mathrm{QNM}]\to 0$.
	
	\textit{Upper temperature limit} To establish a quantitative connection between our holographic model and experimentally accessible parameters in optical lattice BEC systems \cite{64}, we construct a self-consistent mapping between black hole variables and condensate quantities. Specifically, we identify the black hole charge $Q$ with the total particle number $N$ and effective charge $q$ via $Q \sim qN$, the horizon radius $r_h$ with the correlation length $\xi \sim aN^{1/3}$, and the AdS curvature scale $L \sim aN^{1/3}$ with the overall system size, where $a$ is the lattice constant. Substituting these relations into the black hole temperature formula and incorporating the dynamical contribution from many-body cooperative tunneling, we derive an effective expression for the critical temperature of the time-crystalline phase
	\begin{equation}
		T_c\approx \frac{4zJ^2}{k_BU} \left | \psi _0 \right |^2-\frac{ q^2N }{4\pi a^3},
	\end{equation}
	where $J$ is the single-particle hopping amplitude, $U$ the on-site interaction strength, $z$ the coordination number, and $\left | \psi _0 \right |^2$ denotes the condensate amplitude. The first term arises from coherent many-body tunneling that drives collective oscillations, while the second term represents the electrostatic suppression due to the effective charge density. This expression captures the competition between dynamical symmetry breaking and long-range interactions, and provides a direct, testable prediction for the onset of continuous time-crystalline order in strongly coupled lattice condensates. We also derived the holographic duality equations for the spin system and determined its critical temperature $T_c$ 
	
	\textit{Second quantization of time crystal phase} As shown in Fig.1, under sufficiently strong interactions, many-body cooperative tunneling dominates spontaneous symmetry breaking, leading to a reformulation of the Hamiltonian in the Hubbard model (charged black hole) as
	
		\begin{figure}[H]
		\centering
		\includegraphics[width=0.5\textwidth]{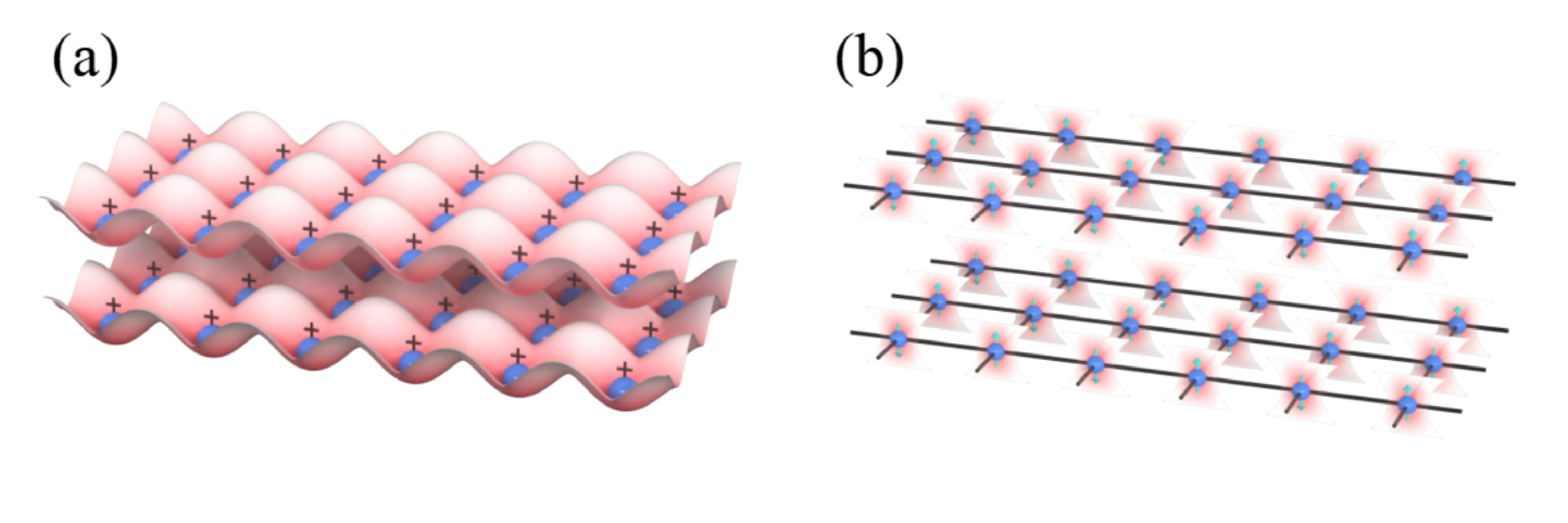}
		\caption{(a) Shows the experimental system of charged particles in a 3D optical lattice. Ionized cold atoms (e.g., Rb87+) undergo many-body correlated tunneling under deep potential wells, causing the atomic number density at lattice sites to oscillate as standing density waves, spontaneously generating a time-crystalline phase. (b) Demonstrates a spin-chain model controlled by optical tweezers. The tweezer manipulation enables long-range phase correlations between spins at different lattice sites, creating spin-density coupling that produces a spontaneous time- crystalline phase.
		}
		\label{fig:rt}
	\end{figure}
	
	\begin{equation}
		\hat{H}_Q =-\sum_{\left \langle i,j \right \rangle} (J \hat{b}_i^\dagger \hat{b}_j+K{\hat{b}_i^{\dagger2}} {\hat{b}_j}^2+\mathrm{h.c.} )+\frac{U}{2}\sum_{i}^{}\hat{n}_i(\hat{n}_i-1 )   
	\end{equation}
 where $U/J\gg 1$, $K=\frac{J^2}{U}$ is double-body tunneling strength. In a quantum time crystal driven by many-body cooperative tunneling, the spontaneous breaking of time-translation symmetry originates from the periodic collective oscillations between many-body quantum states in the lattice. This phenomenon fundamentally arises from the dynamic coupling between density waves and phase coherence. We define the density oscillation operator as the order parameter of the time crystal
	\begin{equation}
		\hat{O}_{\omega Q} =\frac{1}{\sqrt{N} } \sum_{j}^{N}\hat{n} _je^{-i\frac{\pi}{a}r_j },
	\end{equation}
	
	It meet the conditions
	\begin{equation}
		\left \langle\left [ \hat{O}_\omega (t),\hat{O}_\omega ^\dagger (0)   \right ]   \right \rangle=iAe^{-i\omega _nt}.
	\end{equation}
	In Fock space, the single-site particle number operator satisfies the Heisenberg equation of motion
	\begin{equation}
		i\hbar \frac{d\hat{b}_i}{dt} = [\hat{b}_i, \hat{H}] = -J \sum_{j\in\langle i \rangle} \hat{b}_j - 2K \sum_{j\in\langle i \rangle} \hat{b}_i^{\dagger} \hat{b}_j^2  + U \hat{n}_i \hat{b}_i.
	\end{equation}
	With the field of breaking of time-translation symmetry, $\hat{b}_i \to \psi_0 + \delta \hat{b}_i, \quad \langle \hat{b}_i \rangle = \psi(t) = \psi_0 e^{-i\omega_n t}, \left \langle  {\hat{b}^{\dagger 2}} {\hat{b}^2 }\right \rangle \ne 0,	$ we get
	\begin{equation}
		\hbar \omega_n = 4z K |\psi_0|^2.
	\end{equation}
	We can obtain $\omega _n= q \cdot \left( -\left. \frac{\partial \Phi}{\partial r} \right|_{r_h} r_h \right) = q \cdot \frac{Q}{4\pi r_h}=4zK\left | \psi _0 \right |^2/\hbar $, where $z$ is lattice coordination number. The boson number operator begin to oscillate: $\langle \delta \hat{n}_j(t) \rangle \equiv \langle \hat{n}_j(t) \rangle - \bar{n} \propto \cos(\omega_n t + \phi_j)$ (as shown in Fig.2). Being different with superfluid states, time crystals require the simultaneous breaking of: $\langle \hat{b}_j^\dagger \hat{b}_k \rangle \sim e^{-|r_j-r_k|/\xi} e^{i\theta(t)}$, and $\langle \hat{n}_j \hat{n}_k \rangle - \bar{n}^2 \sim \cos[\mathbf{q} \cdot (\mathbf{r}_j - \mathbf{r}_k)]$. We also use the noise measurement (as shown in Fig.3)

	\begin{figure}[htbp]
		\centering
		\begin{subfigure}[b]{0.55\textwidth}
			\includegraphics[width=\textwidth]{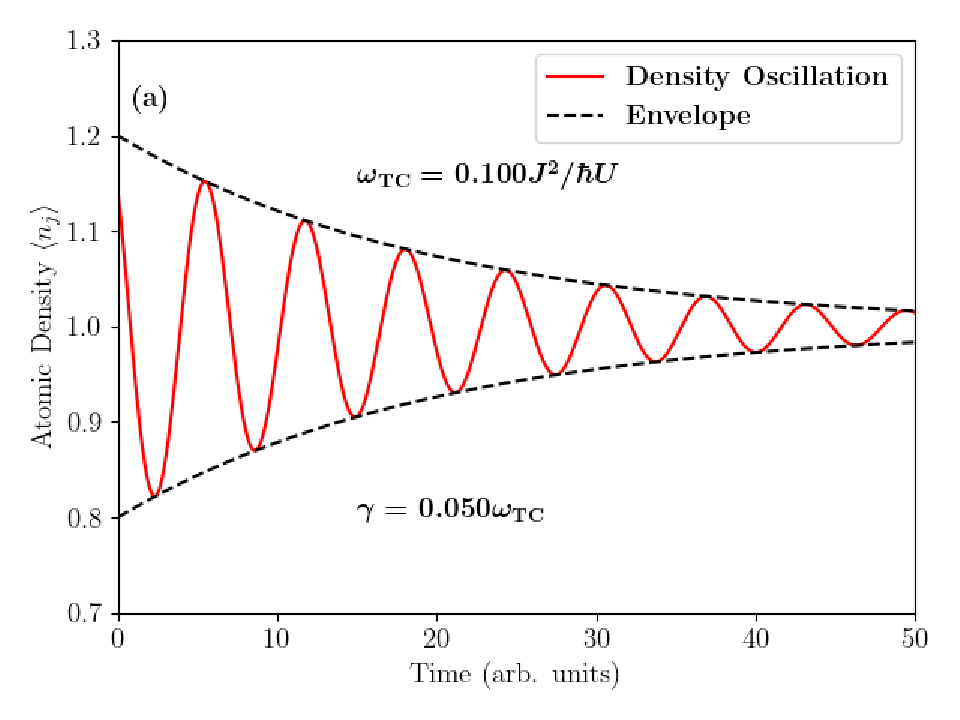}
		\end{subfigure}
		\hfill % 水平间距
		\begin{subfigure}[b]{0.55\textwidth}
			\includegraphics[width=\textwidth]{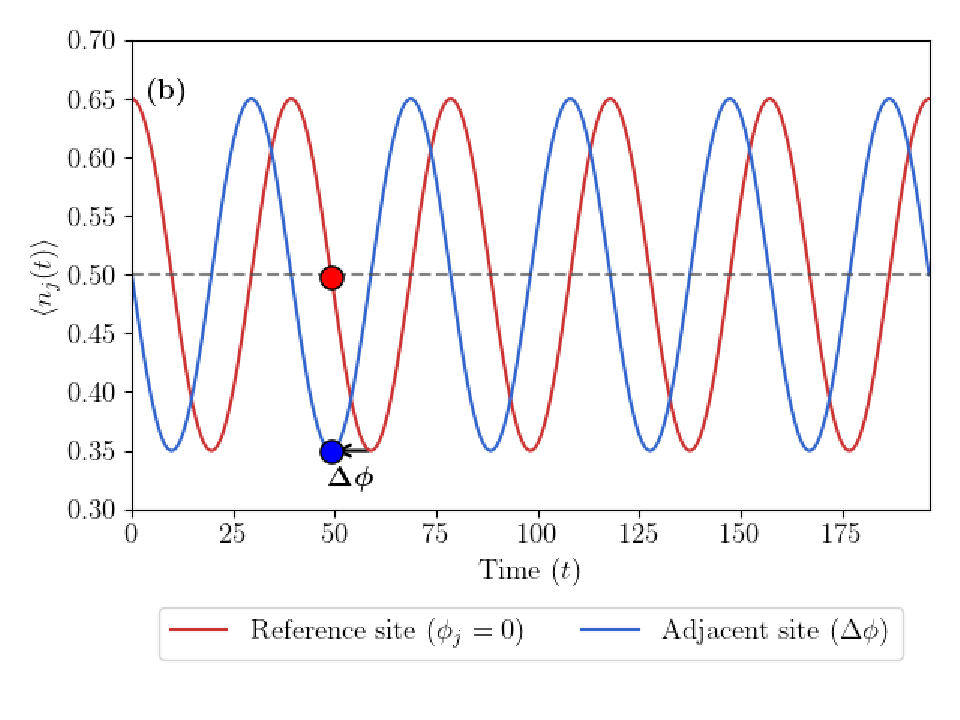}
		\end{subfigure}
		\caption{(a) Time Crystal Density Oscillations ($U/J=10$). In an ideal scenario, the atomic number density of a time crystal exhibits damped oscillations due to background noise. (b) Spontaneous Number Density Oscillation in Time Crystal. In a spontaneously and persistently oscillating time crystal, the atomic number density at different lattice sites differs by a phase shift ($\Delta\phi$), yet the system maintains non-equilibrium stable oscillations without decay. }
	\end{figure}
	
	\begin{figure}[tp]
		\centering
		\includegraphics[width=0.5\textwidth]{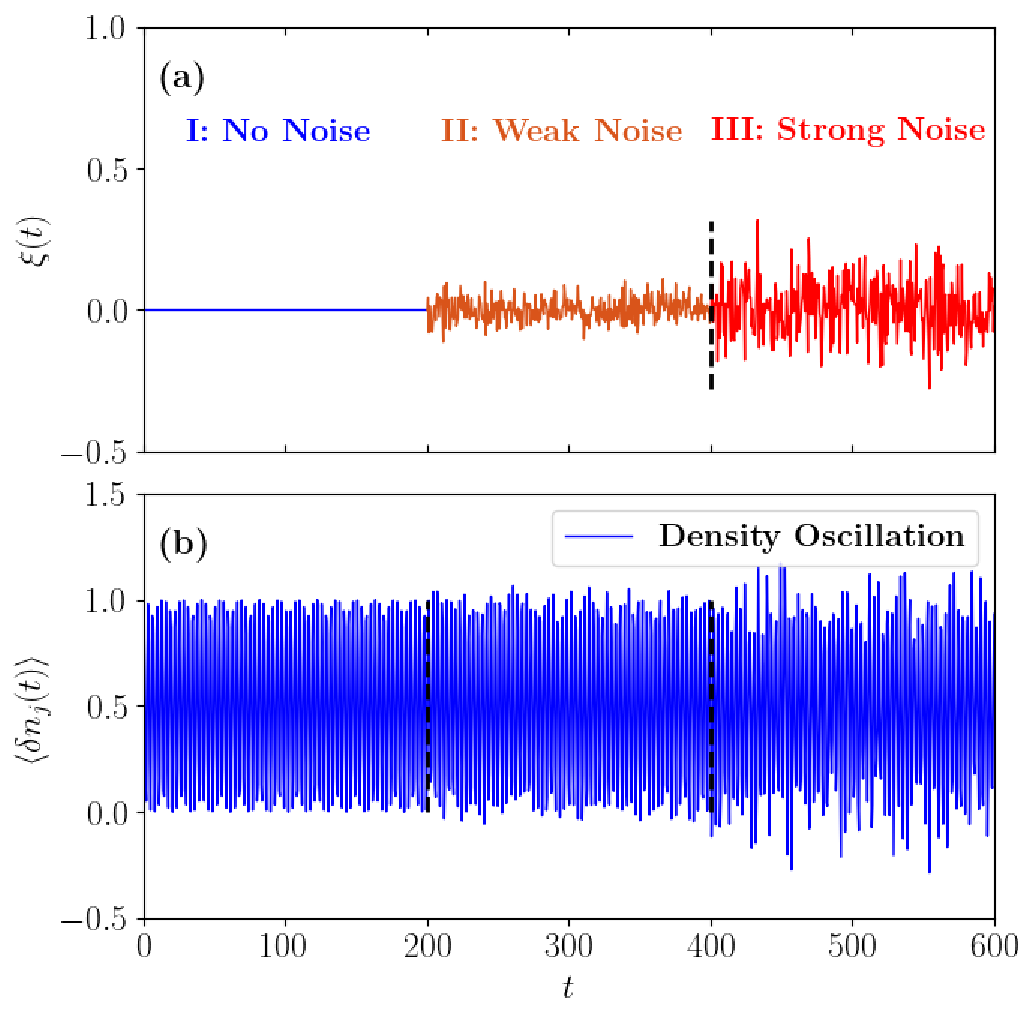}
		\caption{(a) shows 3 kinds of noise:no noise, weak noise and strong noise. (b) shows the density oscillation curves of the system under different noise effects.
		}
		\label{fig:rt}
	\end{figure}
	
	The Heisenberg equation reveals that many-body cooperative tunneling induces eigenoscillations in the density operator, which spontaneously breaks time-translation symmetry via off-diagonal long-range order (ODLRO). It leads to the emergency of a time-crystalline phase, distinct from both superfluid and Mott insulating phases. We get the phase diagram as shown in Fig.4.

	 To ensure consistency between the quasinormal modes of black holes and the sustained oscillations of time crystals, we introduce charged black holes and Kerr black holes. This compensates for the gravitational redshift caused by the Schwarzschild-AdS black hole's decay factor through electrostatic forces, where the black hole's charge corresponds to the chemical potential in the time-crystal system. The relation \(\omega = q\Phi\) guarantees synchronized frequency locking between the time crystal's oscillations and density evolution. Density evolution refers microscopically to periodic oscillations in particle number density (\(\langle \delta \hat{n}_j(t) \rangle\)), while macroscopically manifesting as coupled dynamics between quantum phase and density fluctuations. Within the AdS/CFT framework, the oscillatory dynamics of a continuous time crystal on the boundary are holographically dual to quasinormal modes of a charged black hole in the bulk. The electrostatic potential at the horizon plays a central role in this correspondence: it sets the effective charge for the boundary theory and serves as a phase-locking reference that stabilizes the oscillation frequency of the time crystal. Specifically, the synchronization condition \(\omega = q\Phi(r_h)\) implies that the charged scalar field oscillates at a frequency precisely determined by the black hole’s electric potential, ensuring frequency locking between bulk and boundary dynamics. Physically, this reflects a deeper mechanism where the black hole’s electrostatic field compensates for the gravitational redshift and enables a sustained, dissipationless mode of oscillation at the boundary. Analogous to the Josephson effect, in which a potential difference gives rise to a coherent frequency response, the horizon potential acts as a dynamical energy scale selecting the collective oscillation frequency in the time-crystalline phase. This mapping not only ensures thermodynamic consistency between the two descriptions but also reveals that the emergent time-translation symmetry breaking is fundamentally tied to the holographic duality between charge transport and coherent oscillations. 

	Through holographic correspondence with the scalar field amplitude \(|\psi(r)|\) and charge \(q\) in an RN-AdS black hole,  this evolutionary process is rigorously mapped to  \begin{figure}[H]
		\centering
		\includegraphics[width=0.5\textwidth]{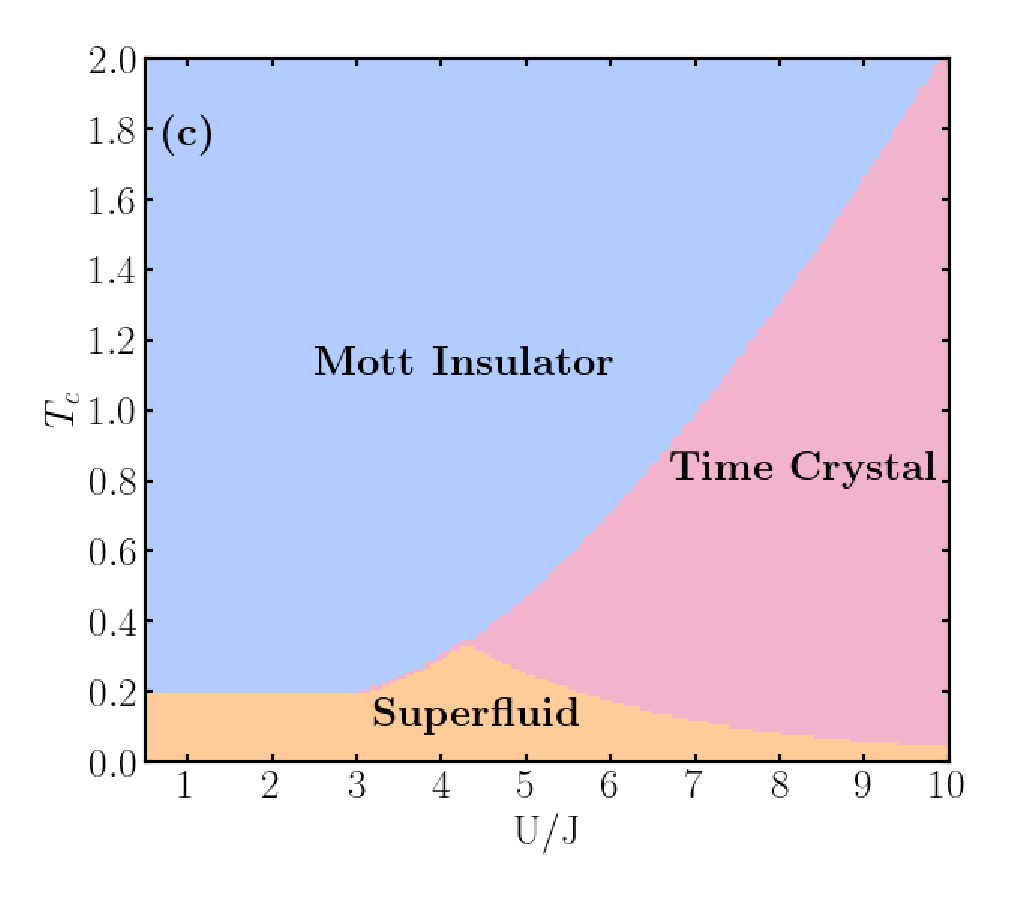}
		\caption{The phase diagram reflects the evolution of the cold atomic system with variations in $U/J$ (interaction-to-tunneling ratio) and temperature. In the low-temperature, strongly correlated regime ($U/J \gg 1$), the time-crystalline phase manifests prominently.
		}
		\label{fig:rt}
	\end{figure}
	\noindent gravitational theory, unveiling the deep mechanism behind the time crystal's sustained oscillations—specifically, the horizon's electric potential \(\Phi(r_h)\) locks the boundary system's oscillation frequency via the synchronization condition \(\omega = q\Phi(r_h)\), achieving dynamic equilibrium between density evolution and phase gradients. At the critical temperature \(T_c\), we find that spontaneous breaking of time-translation symmetry (time crystal) and $U(1)$ symmetry (charge condensation) occur simultaneously. It's a new kind of phase translation. 
	
	\textit{Summary and discussion}.
	In this letter, we employ holographic duality to address time crystals in strongly correlated systems (e.g., 3D optical lattice BECs). The oscillating quantum field of the time crystal is dual to an oscillating charged black hole and Kerr black hole solution in AdS space, with the black hole’s temperature limit mapped to the time crystal’s temperature.  We further demonstrate how many-body correlated tunneling dominates in the strongly correlated Hamiltonian, inducing particle density oscillations—thereby proving the existence of continuous time crystals in strongly correlated regimes via high-order perturbative expansion.

		\pagebreak
	\widetext
	\include{supplemental_materials}

\end{document}

%% file: supplemental_materials.tex
\begin{center}
\textbf{\large Supplemental Materials of Spontaneously Strongly Coupled Holographic Time Crystal}
\end{center}

\setcounter{equation}{0}
\setcounter{figure}{0}
\setcounter{table}{0}
\setcounter{section}{0}
\setcounter{page}{1}
\makeatletter
\renewcommand{\theequation}{S\arabic{equation}}
\renewcommand{\thefigure}{S\arabic{figure}}
\renewcommand{\bibnumfmt}[1]{[S#1]}
\renewcommand{\thesection}{\arabic{section}}
\renewcommand{\thesubsection}{\arabic{section}.\arabic{subsection}}

	\section{Limit temperature $T_c$ of time crystal}
	The time crystal phase corresponds to a conformal field on the boundary of a CFT, where its spontaneous symmetry breaking of time translation is dual to an oscillating charged black hole solution in AdS space. The thermodynamic limit of the black hole is equivalent to the temperature of the time crystal on the CFT boundary. 
	From the RN-AdS metric
	\begin{equation}
	f(r) = 1 + \frac{r^2}{L^2} - \frac{\mu}{r} + \frac{Q^2}{r^2},  \mu = \frac{8\pi G M}{3}. 
	\end{equation}
	Horizon radius $r_h$ satisfies  $f(r_h) = 0$
\begin{equation}
	\mu = r_h \left( 1 + \frac{r_h^2}{L^2} + \frac{Q^2}{r_h^2} \right).
\end{equation}
	The temperature of a black hole is defined by surface gravity
	\begin{equation}
		T = \frac{|f'(r_h)|}{4\pi}, \quad f'(r) = \frac{2r}{L^2} + \frac{\mu}{r^2} - \frac{2Q^2}{r^3}.
	\end{equation} 
	Substitute $\mu$ and simplify
	\begin{equation}
		 T = \frac{1}{4\pi} \left| \frac{2r_h}{L^2} + \frac{1}{r_h^2} \left[ r_h \left(1 + \frac{r_h^2}{L^2} + \frac{Q^2}{r_h^2}\right) \right] - \frac{2Q^2}{r_h^3} \right| = \frac{1}{4\pi} \left( \frac{3r_h}{L^2} - \frac{Q^2}{r_h^3} \right),
	\end{equation}
	where $1/r_h \sim 0$.
	\section{Scaling Law Derivation }
	
	• Scalar mass to interaction
	\begin{equation}
		m L = \alpha \sqrt{\frac{U}{J}}. 
	\end{equation}
	
	• Horizon radius scaling
	\begin{equation}
		r_h \approx \beta (m L^3)^{1/3} = \beta \alpha \left( \frac{U}{J} \right)^{1/3} L. 
	\end{equation}
	
	• Charge to chemical potential
	\begin{equation}
		Q = \gamma \sqrt{\frac{\hbar U}{J}}. 
	\end{equation}
	The correspondence $Q = \gamma \sqrt{\frac{\hbar U}{J}}$ arises from dimensional analysis and scaling arguments in the strongly interacting Bose-Hubbard model, where $Q$ represents a characteristic quantum fluctuation scale (e.g., phonon momentum or phase slip amplitude). Here, $\gamma$ contains a unit of length, while $\sqrt{\hbar U/J}$ combines the competing energy scales: the on-site interaction $U$ (localization energy) and tunneling $J$ (delocalization energy). The ratio  $U/J$  controls the system’s proximity to the superfluid-Mott insulator transition, and the Planck constant $\hbar$ ensures quantum-mechanical scaling. This form is rigorously derived from the model’s low-energy effective theory, where $Q$ quantifies the crossover between particle-like ( $ U$ -dominated) and wave-like ( $J$ -dominated) regimes.
	
	Substitute into $T_c$ 
	\begin{equation}
		k_B T_c = \frac{1}{4\pi} \left( \frac{3}{L} r_h - \frac{Q^2}{r_h^3} \right). 
	\end{equation}
	\section{Perturbation Expanding of Many-body Cooperative Tunneling}
	The original Hamiltonian
	\begin{equation}
		\hat{H} = \hat{H}_0 + \hat{H}_t,
	\end{equation}
	where Unperturbed part is  
	\[
	\hat{H}_0 = \frac{U}{2} \sum_i \hat{n}_i (\hat{n}_i - 1) \quad \text{(On-site repulsion, dominant energy scale } U).
	\]  
Perturbation  
\begin{equation}
	\hat{H}_t = -J \sum_{\langle i,j \rangle} (\hat{b}_i^\dagger \hat{b}_j + \text{h.c.}),
\end{equation}
	(Single-particle tunneling, energy scale $J \ll U$).
 In the strongly correlated limit \( U/J \gg 1 \), derive the effective Hamiltonian up to \( \mathcal{O}(J^2) \) via perturbation theory. A many-body correlated tunneling term \( K \hat{b}_i^{\dagger 2} \hat{b}_j^2 \) should emerge, with \( K = J^2 / U \).
	
\subsection{Schrieffer-Wolff Transformation Framework}
	
	The Schrieffer-Wolff transformation projects the Hamiltonian onto the low-energy subspace via a unitary operator
	 \( e^S \) (where \( S^\dagger = -S \))
	\begin{equation}
		\hat{H}_{\text{eff}} = e^S \hat{H} e^{-S} = \hat{H}_0 + \hat{H}_t + [S, \hat{H}_0] + [S, \hat{H}_t] + \frac{1}{2} [S, [S, \hat{H}_0]] + \cdots.
	\end{equation}
	The first-order term is eliminated by requiring 
	 \begin{equation}
	 	[S, \hat{H}_0] = -\hat{H}_t,
	 \end{equation}
	yielding the effective Hamiltonian
	\begin{equation}
		\hat{H}_{\text{eff}} = \hat{H}_0 + \frac{1}{2} [S, \hat{H}_t] + \mathcal{O}(J^3).
	\end{equation}

	\subsection{Solving for the Generator \( S \)}
	
	From \( [S, \hat{H}_0] = -\hat{H}_t \), \( S \) can be determined. Consider a two-site process \( \langle i,j \rangle \),  
Initial state: \( |n_i, n_j\rangle \) (Fock basis).  
	Action of \( \hat{H}_t \)
 \begin{equation}
 		\hat{H}_t |n_i, n_j\rangle = -J \sqrt{n_i (n_j + 1)} |n_i - 1, n_j + 1\rangle - J \sqrt{(n_i + 1) n_j} |n_i + 1, n_j - 1\rangle.
 \end{equation}
Consider the matrix element condition
	\[ 
	\langle \psi_f | [S, \hat{H}_0] | \psi_i \rangle = - \langle \psi_f | \hat{H}_t | \psi_i \rangle,
	\]  
	where \( \hat{H}_0 \) is diagonal in the Fock basis
	\[
	E_0 = \frac{U}{2} [n_i(n_i - 1) + n_j(n_j - 1)].
	\]  
	For the transition \( |n_i, n_j\rangle \to |n_i - 1, n_j + 1\rangle \),  
	initial energy: \( E_i = \frac{U}{2} [n_i(n_i - 1) + n_j(n_j - 1)] \),  
	and final energy: \( E_f = \frac{U}{2} [(n_i - 1)(n_i - 2) + (n_j + 1)n_j] \),  
	so energy difference: \( \Delta E = E_f - E_i = U (n_j - n_i + 1) \).  
	Substituting \( \hat{H}_t \)’s matrix element
	 \begin{equation}
	 		\langle n_i - 1, n_j + 1 | \hat{H}_t | n_i, n_j \rangle = -J \sqrt{n_i (n_j + 1)},
	 \end{equation}
	we solve for \( S \) 
	\begin{equation}
			{\langle n_i - 1, n_j + 1 | S | n_i, n_j \rangle = \frac{J \sqrt{n_i (n_j + 1)}}{U (n_j - n_i + 1)}}.
	\end{equation}

	\subsection{Calculating the Effective Hamiltonian (Second-Order Term)}

	The key term in the effective Hamiltonian is $
	\hat{H}_{\text{eff}}^{(2)} = \frac{1}{2} [S, \hat{H}_t].$ We focus on two-operator correlated processes (e.g., \( \hat{b}_i^{\dagger 2} \hat{b}_j^2 \)). For the transition \( |n_i, n_j\rangle \to |n_i - 2, n_j + 2\rangle \), there are two possible paths:  
	1. : Tunnel \( i \to j \) to \( |n_i - 1, n_j + 1\rangle \), then tunnel \( i \to j \) again.  
	2. : Tunnel \( j \to i \) (forbidden by particle number conservation).  
	The only viable path involves two consecutive \( i \to j \) tunneling events. The matrix element is
	\begin{equation}
		\langle n_i - 2, n_j + 2 | \hat{H}_{\text{eff}}^{(2)} | n_i, n_j \rangle = \frac{1}{2} \sum_{k} \left[ \langle n_i - 2, n_j + 2 | S | k \rangle \langle k | \hat{H}_t | n_i, n_j \rangle - \langle n_i - 2, n_j + 2 | \hat{H}_t | k \rangle \langle k | S | n_i, n_j \rangle \right],
	\end{equation}
	where the intermediate state \( |k\rangle \) is \( |n_i - 1, n_j + 1\rangle \).  
	Combining terms and simplifying
\begin{align}
		\langle n_i - 2, n_j + 2 | \hat{H}_{\text{eff}}^{(2)} | n_i, n_j \rangle = & \, \frac{J^2}{2U} \sqrt{(n_i - 1) n_i (n_j + 1) (n_j + 2)} \left[ -\frac{1}{n_j - n_i + 3} + \frac{1}{n_j - n_i + 1} \right].
\end{align}
		
	\subsection{Half-Filling Limit} 
	
	Near half-filling (\( n_i \approx n_j \gg 1 \)), set \( n_i = n_j = n \) 
 \begin{equation}
 	\langle n-2, n+2 | \hat{H}_{\text{eff}}^{(2)} | n, n \rangle \approx \frac{J^2}{3U} n(n-1).
 \end{equation}
	Comparing with the matrix element of \( \hat{b}_i^{\dagger 2} \hat{b}_j^2 \) 
	\begin{equation}
		\langle n-2, n+2 | \hat{b}_j^{\dagger 2} \hat{b}_i^2 | n, n \rangle \approx n(n-1) \quad (\text{for } n \gg 1).
	\end{equation}
	And
	\begin{equation}
		K=\frac{1}{U} \left \langle \mathrm{vac} \right | \hat{b}_j^2\hat{H}_t\frac{1}{E_0-H_0}\hat{H}\hat{b}_i ^2\left |\mathrm{vac}\right\rangle=\frac{J^2}{U},
	\end{equation}
	(The coefficients need to be normalized under mean field conditions.) Here is a sketch map:
	\begin{figure}[H]
		\centering
		\includegraphics[width=0.8\textwidth]{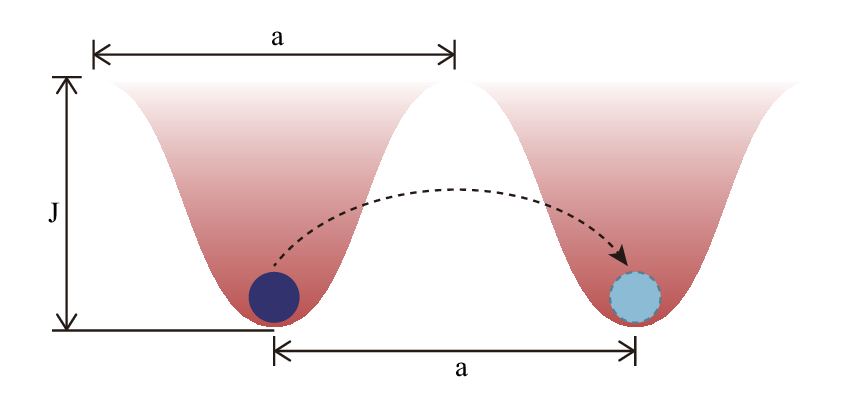}
		\caption{a is the lattice constant.}
		\label{fig:rt}
	\end{figure}
	
	\section{time crystal corresponded to Kerr black hole}
In the many-body system corresponding to spin chain, we still find a correspondence with Kerr black holes in AdS.

In rotating black holes, the frame-dragging effect of Kerr black holes can also counteract gravitational redshift, with the action corresponding to spin-curvature coupling. In ultra-cold atomic systems, the rotating black hole is mapped to particle spin, and the coupling field corresponds to the spinor field. The action reads
\begin{equation}
	S = \int D^{d+1}x \sqrt{-g} \left[ \frac{1}{16\pi G}(R-2\Lambda) - \bar{\Psi} (\gamma^\mu D_\mu - m) \Psi - \lambda (\bar{\Psi} \Psi)^2 \right]
\end{equation} 
where $D_{\mu}=\partial_{\mu}+\frac{1}{4} \omega _{\mu }^{ab}\gamma _{a}\gamma _{b}$ is spin connection, $\gamma^{\mu}$ is Dirac Matrix. The field equation reads:$(i\gamma^\mu D_\mu - m) \Psi = \lambda (\bar{\Psi} \Psi) \Psi.$ And we get the Kerr-metric $ds^2 = -\frac{\Delta_r}{\rho^2} \left( dt - a \sin^2\theta d\phi \right)^2 + \frac{\rho^2}{\Delta_r} dr^2 + \rho^2 d\theta^2 + \frac{\sin^2\theta}{\rho^2} \left( a dt - (r^2 + a^2) d\phi \right)^2.$ We similary obtained $T_c$:
\begin{equation}
	k_BT_c\approx \frac{4zJ^2}{U}\left | \psi _0 \right |^2-\frac{J_{\mathrm{ex}}s^2N}{4\pi a^3},   
\end{equation}
where $J_{\mathrm{ex}}$ is spin-exchange-coupling constant, $s$ is spin quantum number. 

The Hamiltonian in spin model reads:
\begin{equation}
	\hat{H}_s = -\sum_{\langle i,j\rangle} \left( J \hat{b}_{i\sigma}^\dagger \hat{b}_{j\sigma} + J_{\text{ex}} \hat{\mathbf{S}}_i \cdot \hat{\mathbf{S}}_j \right) + \frac{U}{2}\sum_i \hat{n}_i(\hat{n}_i-1),
\end{equation}
where $\hat{\mathbf{S}}_i = \frac{\hbar}{2} \sum_{\alpha\beta} \hat{b}_{i\alpha}^\dagger \vec{\sigma}_{\alpha\beta} \hat{b}_{i\beta}$ is spin operator.And we can obtain the order operator as
 \begin{equation}
	\hat{O}_{\omega s} = \frac{1}{\sqrt{N}} \sum_j \hat{n}_{j} e^{-i\frac{\pi}{a} r_j} \otimes \hat{S}_j^z .
\end{equation}

In rotating black holes, frame-dragging effect, arising from spacetime torsion induced by black hole spin, modifies the radial potential barrier: the effective gravitational redshift is suppressed as the dragged spacetime reduces energy loss in radial motion, allowing coherent oscillations near the horizon. Analogously, in cold atomic spin systems, spin-exchange interactions (e.g., Heisenberg model $H_{\text{ex}} = J\sum \mathbf{S}_i \cdot \mathbf{S}_j$) induce long-range spin correlations, forming a "spin-dragged" state. These correlations act like an effective "spin frame-dragging," suppressing single-particle decoherence by redistributing energy dissipation from individual tunneling to collective spin modes. The collective spin oscillations (analogous to black hole's horizon-bound coherent modes) break time-translation symmetry spontaneously, yet their correlated dynamics stabilize the tunneling barrier—reducing it via $V_{\text{eff}} \propto V_0/N ( N$ : correlated spins), similar to how frame-dragging lowers the radial potential near a Kerr black hole. This synergy between spin correlations and collective tunneling enables enhanced coherence and suppressed barrier penetration, bridging black hole gravity and quantum many-body physics.